# An Approach to Learning Research with a Wireless Sensor Network in an Outdoor Setting

**ANDERSON, Tom Adam Frederic[a], WEN, Yean-Fu[b]**
[a]*Graduate Institute of Network Learning Technology, National Central University, Taiwan*
[b]*Department of Information Management, Chinese Culture University, Taiwan*
ta@cl.ncu.edu.tw

Abstract: Automated collection of environmental data may be accomplished with wireless sensor networks (WSNs). In this paper, a general discussion of WSNs is given for the gathering of data for educational research. WSNs have the capability to enhance the scope of a researcher to include multiple streams of data: environmental, location, cyberdata, video, and RFID. The location of data stored in a database can allow reconstruction of the learning activity for the evaluation of significance at a later time. A brief overview of the technology forms the basis of an exploration of a setting used for outdoor learning.

## Introduction

A pervasive information-gathering technology comprised of wirelessly connected nodes with remote sensing and computing capabilities, wireless sensor networks (WSNs) are fundamental to hundreds of applications in research and industry. For more general information on WSNs, we recommend the work of [2] and [11], indispensible in the preparation of this work. This paper aims at a general data-gathering system for outdoor learning.

A wireless sensor transmits information about aspects of the environment that it senses. In educational settings, wireless RFID enable wireless identification and location of radio frequency tags, with research applications in learning, such as language learning [7] yet not without challenges [8]. With more robust WSNs, researchers have enabled the creation of kindergarten objects that are responsive to stimulus [12], and for capturing many streams of information from lecturers and students in university classes [1]. Although many challenges are involved in using WSNs for educational research, WSNs afford new capabilities to multidisciplinary research groups including educational researchers.

## 1. Behavior Settings

The behaviors of participants are *synomorphic* with, or similar to, the setting, which includes locations, physical and temporal attributes and other participants [10]. Elements of the environmental are implicit in learning activities, but the scope of manual data collection techniques is limited in focus. Just as the evaluation of exhibits and visitor studies can be enhanced by automated data gathering in a museum [14], WSNs can aid in the monitoring of the behaviors of learners and other participants in outdoor settings.

We address the specific purpose of investigating an outdoor mobile learning activity. In such a location, learning researchers typically use field study techniques such as participant observation, interviews, and handheld video recording. Usage of devices for data collection should be a carefully planned part of research design, and techniques should be piloted to ensure success in the field [9].

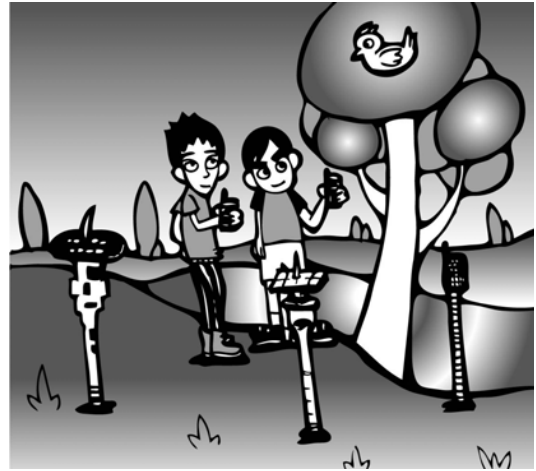

The proposed site for deployment is an outdoor area. Established for recreational purposes and including a pond, trees, and abundant wildlife, this nature area is a popular destination on weekends and holidays; however, during the workweek it is underused. Teachers can direct educational day trips during the week when the park is less populated, but to better study the education value of such visits, the WSN will allow researchers to unobtrusively observe activities, as shown in Figure 1, in order to improve the understanding of the activities that take place.

**Fig. 1. Learner activity remotely monitored by three wireless sensors, shown in foreground.**

## 2. Wireless sensor network design

The feasibility of constructing a wireless multimedia sensor network [2] for outdoor research is primarily constrained by system cost, battery lifetime, and bandwidth considerations, but allow the automated collection and tagging of the following data types:

- *Environmental factors* – This data, including temperature, humidity, ambient audio, population density [Lewis'04], can be assessed for impact on learning.
- *Cyberdata* – The recording of devices used by learners. The design of mobile learning activities, such as bird-watching expedition using PDAs [4], and studying plants [3], could potentially be enhanced by collecting cyberdata, location-tagged for analysis. Audio can also be collected through on-person devices.
- *Video sensor network* – In this work, we adopt low-power/solar-powered 360-degree video sensors as introduced by [6], suitable for remote deployment. Nodes periodically wake to compare the field of view with prior images. If there are no changes, a node will enter sleep mode to conserve energy. Instructions to wake are transmitted to neighboring nodes when movements are sensed.
- *RFID* – An RFID scanner is capable of collecting identification and location of tagged people and objects. This information will subsequently be used for associating streams of environmental, cyber, and video data in the database.

Learners and teachers may also profit from environmental, cyber-, video and RFID data. Due to energy considerations, communications between wireless sensors are connected by low-powered frequencies, and higher bandwidths video feeds, are sent over 802.11a/b/g. Environment sensing enhances automatic labeling, tracking movements and conditions.

*2.1 Deployment and data gathering*

The primary goal of sensor deployment is the maximum coverage of the area of interest within constraints of sensor life and transmission radius. Sensor deployment is divided into two major approaches, pre- and post-deployment. Pre-deployment is appropriate for video sensor networks and RFID sensor networks that require good point of view to identify objects. Accordingly, sensor nodes will be placed so as to provide maximum coverage of the area of interest, such that all participants and items of interest are captured. Post-deployment WSN communications are self-configured; sensor node data is aggregated and forwarded via location-based routing to the gateway, as in Figure 2. Query-based, negotiation-based, QoS-based routings are adopted, allowing data to be obtained interactively in real-time, enabled by wireless communications techniques addressed in [2] [5] [13].

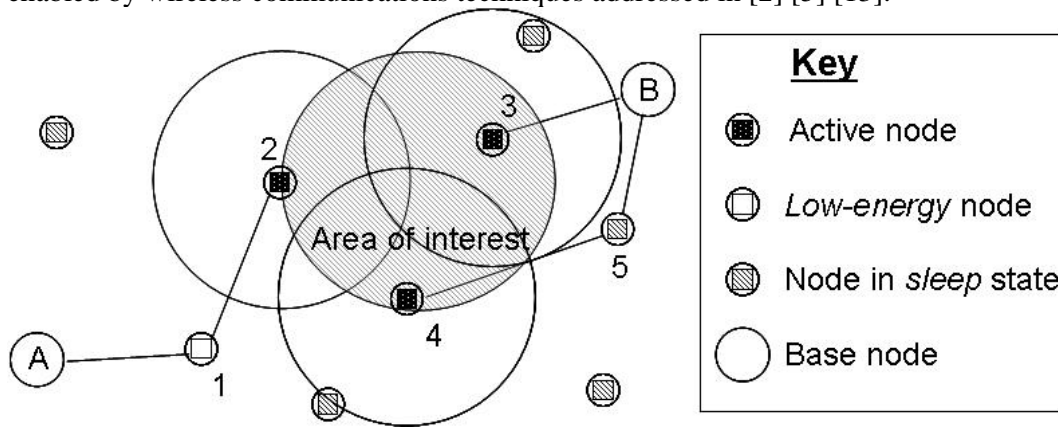

**Fig. 2. Sensor nodes (2, 3, 4) capture data from shaded area of interest, which is then routed to base stations (A, B) over paths [4, 5, B], [3, B], [2, 1, A].**

Adjustments to sensors and cameras are centrally controlled interactively according to application. The transparency of nodes relieves researchers of constant reconfiguring of the network. Cyberdata collected from a mobile learning device such as a PDA may be forwarded on demand for real time analysis, coupled with data from audio streams captured by learner-worn microphones. Audio, video and cyberdata is identified via RFID for future analysis. Recordings of learners and researchers will facilitate subsequent discussions and refinement of findings. Data mining and database theory applying to large datasets allow researchers to determine patterns in existing data, from which theories can be developed.

## 3. Discussion and Implications

Wireless sensor networks allow researchers to collect comprehensive data from wide ranges of time and place. A large database of multiple streams of data will be a great asset for researchers; however, the difficulty of discovering patterns within such a database remains a crucial concern for the future. Future work will focus on improving fault tolerance and quality, maximizing the coverage and lifetime of the network while minimizing costs, and better meeting the needs of researchers. We aim to demonstrate that environmental factors such as weather and learner density can be connected with learning. We anticipate potential expansions including location-based real-time surveys using mobile devices, and the gathering of biomedical data from the learners such as heart rate, and so on. Identifying the suitability of the data for the needs of the researchers remains a vital concern. In the long term, we seek to develop ways that real-time data can be used to create environments that actively react to events.

*3.1 Conclusion*

Wireless sensor networks have found widespread utility in many research domains. WSNs will be found to provide benefit to educational research in the coming decade and beyond. The contribution of this paper is to provide a basic overview of a WNS system designed for automated data gathering in an outdoor learning setting, so as to determine relationships between environmental features and observable behaviors of learners. The work would have multifold purposes, including: to gather data that influences educational theory; to develop concepts of how educational researchers can use WSNs to enhance data gathering abilities within bounds of affordability and efficiency; to expand knowledge of wireless multimedia sensor networks, such as battery life, routing and topology.

The purpose of collecting many different streams of data is not for the simple purpose of generating data, but rather, these tools will be used deliberately for the purposes of establishing significant phenomena in scales that are otherwise too difficult, too expensive, or too time intensive. Further investigations will probe the use of WSNs to enrich learning experiences by allowing teachers and learners to use the capabilities of WSNs and by facilitating greater degrees of interactivity in learning environments.